\documentclass[12pt]{article}
\usepackage[]{graphicx}
\usepackage[]{epsfig}

\newcommand {\be} {\begin{equation}}
\newcommand {\ee} {\end{equation}}
\newcommand {\bea} {\begin{eqnarray} }
\newcommand {\eea} {\nonumber \end{eqnarray}}
\newcommand {\eps} {\epsilon}
 \newcommand {\si} {\sigma}

\newcommand {\ba} {\overline}
\newcommand {\lan} {\langle}
\newcommand {\ran} {\rangle}

\newcommand {\cB}  {{\cal B}}
\newcommand {\cC}  {{\cal C}}

\newcommand {\cP}  {{\cal P}}

\newcommand {\cT}  {{\cal T}}

\newcommand {\bc} {\begin{center}}
\newcommand {\ec} {\end{center}}
\newcommand {\bd}{\begin{displaymath}}
\newcommand {\ed}{\end{displaymath}}

\newcommand {\tgh} {\mbox{th}}
\newcommand {\arth} {\mbox{arth}}

\newcommand {\for} {\ \ \ \mbox{for}\ \ }

\def \form#1 {eq. (\ref{#1}) }
\def \parziale#1#2  {{\partial {#1} \over \partial {#2}}}

\title{Constraint Optimization and Statistical Mechanics}

\author{Giorgio Parisi  \\
Dipartimento di Fisica, INFM, SMC and INFN, \\
Universit\`a di Roma {\em La Sapienza}, P. A. Moro 2, 00185 Rome, Italy. }

\begin{document}
\maketitle
\begin{abstract}
In these  lectures I will present an introduction to the results that have been recently obtained in
constraint optimization of random problems using statistical mechanics techniques.  After presenting
the general results, in order to simplify the presentation I will describe in details only the problems
related to the coloring of a random graph.
\end{abstract}
\section{Introduction}
Statistical mechanics of disorder systems and constraint optimization of random
problems have many point in common.  In these lectures I will describe some of the results that have
been obtained by applying the techniques developed in the study of disordered systems to optimization
theory.  The aim of these studies is twofold:
\begin{itemize}
    \item Statistical mechanics techniques are very sophisticated and powerful: using them it is possible 
    to obtain very relevant heuristic and eventually exact results for optimization theory. Better 
    algorithms can also be found.
    \item Applying the  techniques of statistical mechanics in a rather different setting needs the 
    invention of new theoretical tools that can be carried back to the study of systems of physical 
    interest.
\end{itemize}

This cross fertilization process between  different fields  usually produces very interesting results.

These lectures are organized as follows.  In the next section I will present some general
considerations on the relation between optimization theory and statistical mechanics.  In section III I
will describe a well understood problem, i.e. bipartite matching, for which many analytic and numeric
results are available.  In the next section I will introduces some of the basic ideas and notations
in constraint optimization and I will stress the relations with statistical mechanics and with the theory of
phase transitions. In section V I will recall the main properties of random lattices and of the
Bethe (cavity) approximation, that will be heavily used in the rest of these lectures. In the next 
section I will present the problem of graph coloring and I will derive the appropriate cavity equations 
in the colorable phase. In section VII I will sketch the survey method that allows us to compute in
an (hopefully) exact way the phase transition between the colorable and the uncolorable phase.
Finally in the last section I will present some conclusions and perspectives.

\section{General considerations}
In statistical mechanics \cite{I} one is interested to compute the partition function as function of the
temperature, the partition function being defined as 
\be 
Z(\beta)=\sum_{\cC}\exp( -\beta H(\cC))\ ,
\end{equation} 
where $\beta=1/(kT)$, $T$ being the temperature and $k$ being equal to two thirds of the
Boltzmann-Drude constant.  The variable $\cC$ denotes a generic element of the configuration space.
The quantity $H(\cC)$ is the Hamiltonian of the system.  Starting from $Z(\beta) $ we can reconstruct
the other thermodynamic quantities.
 
In optimization problems we are interested to find out the configuration $\cC^{*}$ (or the
configurations, as far as the ground state may be degenerate) that minimizes the function $H(\cC)$,
that in this contest is the called the cost function. We are also interested in knowing  the minimal
cost, i.e.
\be
 H^{*}\equiv H(\cC^{*}) \ .
 \ee

It is trivial to realize that optimization is a particular case of statistical mechanics \cite{II}.
Let us denote by $E(T)$ and $S(T)$ the expectation value of the energy and the entropy as function of
the temperature $T$.  One finds that only the zero temperature behaviour is relevant for
optimization: 
\be 
H^{*}=E(0) 
\ee 
and the number of optimizing configurations is just equal to $\exp(S(0))$.
 
Often in optimization we are interested in knowing something more (e.g. their number) about the
nearly optimizing configurations, i.e. those configurations such that $H(\cC)=H^{*}+\eps$.  This
information may be gathered by studying the system at small temperatures.
 
In both cases, statistical mechanics and optimization theory, we are interested to study what happens
in the thermodynamic limit, i.e. when the number $N$ of variables (the dimensions of the
configuration space) goes to infinity, or it is very large (finite size effects are more important in
optimization theory than in statistical mechanics).

While optimization theory could be considered only from an abstract point of view as the zero
temperature limit of statistical mechanics, in reality it differs from statistical mechanics in many
crucial aspects, e.g. goals, problems and techniques.  Optimization theory is really a different
science, with many points of contact with statistical mechanics.

Optimization problems, where the Hamiltonian has a simple form, are not usually the mostly
interesting (on the contrary an incredible large amount of work has been done in statistical
mechanics on the Ising problems).  In optimization theory it is natural to consider an ensemble of
Hamiltonians and to find out the properties of a generic Hamiltonian that belongs to this ensemble.

In other words we have an Hamiltonian that is characterized by a set of parameters (collectively
denoted by $J$) and we have a probability distribution $\mu(J)$ on this parameter space.  One would
like to compute the ensemble average\footnote{Sometimes the ensemble is defined in a loose way, e,g.
problems that arise from practical instance, e.g. chips placements on a computer board.  or register
allocation}, e.g.
\be
E_{AV}(T)=\int d\mu (J) E_{J}(T) \equiv \overline{E_{J}(T)} \ ,
\ee
where the overline denote the average of the ensemble of Hamiltonians.  Eventually we are going to set
$T=0$.  This approach has been developed in statistical mechanics for studying systems with quenched
disordered (e.g. spin glasses) \footnote{Many of the techniques developed for disordered systems can be used
also for the study of structural disordered systems where no disorder is present in the Hamiltonian.}.
Only the recent development of these techniques allow us to use statistical mechanics tools to study
optimization theory.

In general we are interested in computing the probability distribution $P(E)$ of the zero
temperature energy $E$ over the ensemble:
\be
P(E)=\overline{\delta(E_{J}(0)-E) } \ .
\ee
In the thermodynamics limit, where the number $N$  of variables goes to infinity, if $E$ 
is well normalized in this limit, we expect that its probability distribution becomes a delta function, 
according to the general principle that its 
intensive quantity do not fluctuate in the thermodynamic limit.

The interest of computer scientists is not limited to the computation of the ensemble average: they
live in the real world and they are extremely interested to find an efficient algorithms to compute
for a given $H_{J}(\cC)$ (i.e. for an instance of the problem) the configuration $\cC^{*}$ that
minimizes it.  Of course the configuration $\cC^{*}$ can always be found (at least in the case of a
finite configuration space) by computing $H(\cC)$ for all possible choice of the configurations, but
this algorithm is very very slow \footnote{This algorithm usually takes a time proportional to
$\exp(A N)$.}.  One would like to find out the most efficient algorithms and to correlate their
performances to the properties of the ensemble of problems.

A given algorithm can be tested by checking how it performs for different instances: for each problem
there are well known benchmarks that correspond to different types of ensembles.  We can also do a
more theoretical analysis.  For a given algorithm we can define the time $t_{J}$ as the time (i.e.
the number of operations) it takes to find the ground state of the Hamiltonian $H_{J}$.  We can
define the average time (in a logarithmic scale) as: \be
 \ln (t_{ln})= \overline{\ln (t_{J})} \ .
 \ee
The introduction of the logarithm is very important: very often the linearly averaged time
\be
t_{L}= \overline{ t_{J}} 
\ee
is much greater than $t_{AV}$ because it may be dominated by rare configurations that give an extremely 
large contribution \footnote {This is the same argument that implies that the annealed free energy and 
the quenched free energy are very different: $\overline{\ln(Z_{J})}$ is very different from 
$\ln(\overline{Z_{J}})$.  }.  We can also define the most likely time, ($t_{ML}$) as the time where the 
probability distribution of the time has a maximum and the median time. 

In many cases the logarithmically averaged time, the median and the most likely times behave in a quite 
similar way when the number $N$ of variables becomes large: in the best of the possible worlds
\be
\ln (t_{J})/\overline{\ln (t_{J})} \ 
\ee
does not fluctuate in the thermodynamics limit.

On the contrary  very often the worst case time, defined as
 \be
 t_{WC}=\max_{J}(t_{J}) \ ,
  \ee
is much larger that the other times \cite{GaJo}.  
The most likely time (or the logarithmically averaged time) is the 
most interesting, unfortunately the computation of the worst case time is the the most easy one from the 
theoretical point of view (in empirical studies, i,e.  when testing the algorithm with real
instances, the 
situation is just the opposite).  

It is rather difficult to study analytically  the most likely time:  if the 
algorithm is interpreted as a physical system, we must study its time evolution toward equilibrium.  In 
spite of the spectacular progresses that have been done in recent years, a general theory of the 
approach to equilibrium is lacking and this is one of the most difficult problem of statistical 
mechanics. Fortunately present days theory allow us to do some predictions \cite{PARISILH,LET}.

\section{A well understood problem: bipartite matching}  

Let us be specific and let us consider an example of optimization problem for which we have a good 
understanding.  Let us consider the so called bipartite matching.  

The problem is defined as follows: we have $N$ cities and $N$ wells that must be connected one with
the other: in other way there is one and only one well $\pi(i)$ that is connected (matches) to the
city $i$ \footnote{In simple matching we have $N$ cities that must be pair-wise matched: here the
configuration space contains $(N-1)!!$ elements.}.  In other words $\pi(i)$ is a permutations and the
configuration space of all possible matching contains $N!$ elements.
 
The Hamiltonian $H_{d}(\pi)$ is given by 
\be
 H_{d}(\pi) =\sum _{i=1,N} d_{i,\pi(i)} \ , \label{BIP}
 \ee
where $d_{i,k}$ is a  $N \times N$ matrix expressing the cost of establishing a connection from 
$i$ to $k$ an it characterizes an instance of the problem.
 
In spite of the fact that the configuration space contains $N!$ elements there are algorithms that
for any $d$ compute the minimum energy configuration $\cC^{*}(d)$ and the corresponding energy
($E^{*}(d)$) in a time that is less that $N^{3}$ multiplied by an an appropriate constant.  The
algorithmic simplicity of the model is related to the fact that it can be recast as a linear
programming problem.  Indeed we can introduce the $N^{2}$ variables $n_{i,k}=0,1$ such that
\be
\sum_{i=1,N}n_{i,k}=\sum_{k=1,N}n_{i,k}=1 \ . \label{VINCOLO}
\ee
It is obviously that there is an one to one correspondence of these variables with the permutation
and the Hamiltonian may be rewritten as
\be
 H_{d}(n) =\sum_{i,k=1,N} d_{i,k} n_{i,k}\ .
 \ee
Up to this point we have not gained too much.  However we can enlarge the configuration space by
considering real variables $s_{i,k}$ that satisfies both the constraint eq.(\ref{VINCOLO}) (i.e.
$\sum_{i=1,N}s_{i,k}=\sum_{k=1,N}s_{i,k}=1$) and the bound
\be
0\le s_{i,k}\le 1 \ .
\ee
The problem is more general and it can be solved using linear programming tools in a very efficient way. 
The surprise is that the minimum of 
\be
H_{d}(s) =\sum_{i,k=1,N} d_{i,k} s_{i,k}
\ee
happens always when $s_{i,k}=0,1$, as can be easily proved. In this way, by enlarging the configuration 
space, we are able to find a fast solutions of the original problem.

From the algorithmic point of view the situation is quite good.  A different set of questions arise
if we consider the ensemble average of the value of $H$ at the mimima.  In this case we have to
specify the ensemble by specifying the probability distribution of the matrix $d$.

A well studied case is when the matrix elements
$d_{i,k}$ are independent identically distributed (i.i.d.) variables: here we have to assign the 
probability of a single matrix element $P(d_{i,k})$.  Simple results are obtained when
\be
P(d_{i,k})=\exp(-d_{i,k}) \label{P0}
\ee
i.e $d_{i,k}=-\ln(r_{i,k})$, where $r_{i,k}$ is a random number, with flat probability distribution in 
the interval $[0-1]$. 

It was firstly conjectured \cite{CON} an then proved rigorously \cite{RIG}
that 
\be
\overline{H(d)}=\sum_{n=1,N}\frac{1}{n^{2}} \ .
\ee
The proof of this very simple result is a tour de force in combinatorial optimization \cite {RIG} and
it is rather unlikely that similar results are valid for different choices of the function $P(d)$.

The situation becomes simpler if we consider the case where $N$ goes to infinity. Using the techniques of 
statistical mechanics \cite{BIPA0} we can prove that
\be
\lim_{N\to \infty} \overline{H(d)} = \zeta(2)=\frac{\pi^{2}}{6}
\ee
for  all the probability distributions such that $P(0)=1$. More over if 
\be
P(d)=1-Ad +O(d^{2}) \ ,
\ee
one finds that \cite{BIPA0,BIPA1,BIPA2}
\be
\overline{H(d)}=\zeta(2)- \frac{2(1-A)\zeta(3)+1}{N}+O(N^{-2}) \ .
\ee 
This result is in agreement with previous formula for the exponential probability where $A=1$.  The
computation of the leading term when $N\to \infty$ is nowadays a real mathematical theorem
\cite{ALDOUS} (that was obtained 15 years after the original result derived using the usual
non rigorous methods of statistical mechanics), while the computation of the subleading term (the one
proportional to $1/N$) was only done using statistical mechanics techniques.

The computation of the subleading term is a highly non-trivial computations, the first two attempts
gave wrong results (a factor two \cite{BIPA1} and a Jacobian \cite{BIPA2} were forgotten), however
now we have the correct versions \cite{BIPA2}.  It has been verified with at least 3 decimal digits.
The computation of the term $-2(1-A)\zeta(3)/N$ is rather simple.  The real difficulty is the term
$-1/N$ (in the second computation with the missing Jacobian this term was $-\pi^{2}/(12 N)$), however
it correctness can be checked with the exact result for the exponential probability.

A similar problem is the travelling salesman: here we have to find the least expensive tour that
travels through all the cities (cities and wells are identified).  The matrix $d$ represent the cost
for going from city $i$ to $k$ and it usually taken to be symmetric.  The configuration space is
restricted to permutations that contains only one cycle (i.e. cyclic permutations) and the
Hamiltonian has always the same form (i.e. eq.(\ref{BIP})).

It is amusing to note that, when we restrict the configuration space, the nature of the problem changes
dramatically and it became much more difficult.  The travelling salesman problem is $NP$ complete,
this statement implies (if a well known conjecture, worth one million dollars, is correct) that there
is no algorithm that can find a solution of the travelling salesman problem in time that is bounded
by a polynomial in $N$ in the worst case \footnote{NP does not means non-polynomial, but Polynomial
on a Non-standard machine \cite{GaJo}, e.g. a computer that has an unbounded number of nodes that
work in parallel, however it is quite likely that this misinterpretation of the name is not far from
reality.}.  However it is empirically found (and I believe it has also been proved) that, in the
typical case of random distribution of the elements of the matrix $d$, a standard simple algorithm
takes a time that does not increase faster that $N^{4}$.  This example shows that there is a dramatic
difference among the time in the typical case and in the worst case.  Unfortunately while it is
relatively easy to measure the scaling of the typical time as function of $N$, the results for the
worst cases are very difficult to be obtained numerically.

These examples are particularly interesting because they tell us that the scenario may radically
change by an apparently small changes in the definition of the problem.  However from the point of
view of analytic computations the results are only slightly more difficult in the traveling salesman
problem.  If $P(0)$ (defined in eq.  (\ref{P0})) is equal to one, we find that in the large $N$ limit
\be
\overline{H(d)}=2.041.
\ee
Apparently there are no serious difficulties in computing the $1/N$ corrections to the traveling
salesman problem and the $1/N^{2}$ corrections to the bipartite matching, however the computations
become rather long and they have not been done.

Summarizing statistical mechanics techniques, based on the replica approach or the cavity approach, are 
able to provide the exact solution to these models in the thermodynamic limit: they can be used to 
compute  also the leading finite size correction (that is a highly non-trivial test).

These methods are working very well when there is no structure in the ensemble of Hamiltonians (i.e. the 
$d$'s are independent identically distributed variables).  
A  different situation may arise in other cases.  For example let us define a three dimensional
bipartite matching problem: we  consider a cube of size $1$, we  extract randomly $n$ points
$x_{i}$ and $y_{i}$ inside the cube and we construct the matrix $d$ using the Euclidean distance
among the points:
\be
d_{i,k}=|x_{i}-y_{k}| \ .
\ee
In this way the distances are correlated random variables (e.g. they satisfy the triangular
inequality) and the exact computation of the minimal energy state is at least as difficult as the
computation of the free energy of the three dimensional Ising model.  Both problems have a three
dimensional structure that is absent in models where the distances are uncorrelated.

The situation where the distances are uncorrelated plays the role of a mean field theory and the
results in finite dimensions can be approximately obtained by doing a perturbative expansion around
the uncorrelated mean field case \cite{MATD0}.  The results are rather good (there are still some
points that are not so well understood \cite{MATD1}), however the description of these techniques
would lead us too far from the main goal of these lectures.

\section{Constraint optimization}

\subsection{General considerations}
Constraint optimization is a particular case of combinatorial optimizations and it will be the main 
subject  of these lectures.  
Let us consider a simple case: a configuration of our system is composed by a number $N$ of variables
$\sigma_{i}$ that may take $q$ values (e.g. from 1 to $q$). These values usually have some meaning in
real life, but this does not necessarily concern us.  A instance of the problems is characterized by
$M$ function $f_{k}[\sigma]$ ($=1,M$), each function takes only the values 0 or 1 and depends of a
small number of $\sigma$ variables.  

Let us consider the following example with $N=4$ and $M=3$:
\bea
c_{1}[\sigma]=\theta(\si_{1}\si_{2}-\si_{3}\si_{4}) \ , \\
c_{2}[\sigma]=\theta(\si_{1}\si_{3}-\si_{2}\si_{4}) \ ,\nonumber  \\
c_{3}[\sigma]=\theta(\si_{1}\si_{4}-\si_{2}\si_{3})  \ ,
\eea
where the function $\theta(x)$ is 1 if the argument is non-negative and it is 0 if the argument is negative.
The function we want to minimize is 
\be
H[\sigma]=\sum_{k=1,M}c_{k}[\sigma]\ .
\ee

In particular we are interested to know if there is a minimum with $H[\sigma]=0$.  If this happens
all the function $c_{k}$ must be zero.  It is quite evident that imposing the
condition $H[\sigma]=0$ is
equivalent to finding the solution to the following inequalities:
\bea
\si_{3}\si_{4}>\si_{1}\si_{2} \ , \\
\si_{2}\si_{4}>\si_{1}\si_{3} \ , \nonumber  \\
\si_{2}\si_{3}>\si_{1}\si_{4} \ .
\eea
In other words each function imposes a constraint and the function $H$ is zero if all the
constraints are satisfied.  If this happens, the minimal total energy is zero and we are in the
satisfiable case. On the contrary, if not possible to satisfy all the constraints, the
minimal total energy is different from zero and we stay in the unsatisfiable case.  In this case the
minimum value of $H$ is the minimum number of constraints that have to be violated.  It is clear that
for each set of inequalities of the previous kind there are arithmetic techniques to find out if
there is a solution and, if any, to count their number.

\subsection{The thermodynamic limit}
Given $N$ and $M$ we can easily define an ensemble as all the possible different set of $M$ inequalities 
of the type
\be
\si_{i_{1}(k)}\si_{i_{2}(k)} >\si_{i_{3}(k)} \si_{i_{4}(k)}\ .
\ee

The interesting limit is when $N$ goes to infinity with 
\be
M=N\alpha \ , 
\ee
$\alpha$ being a parameter.  Hand waving arguments suggest that for small $\alpha$ it is should be
possible to satisfy all the constraints, while for very large $\alpha$ also in the best case most of
the constraints will be not satisfied. It is believed that in this and in many other similar problems
there is a phase transition from a satisfiable to a non-satisfiable phase \cite{sat0}.  More
precisely let us define the energy density
\be
e(\alpha,N)=\frac{\overline{H^{*}}}{N}
\ee
where the average is done over all possible constraints over the set of $N$ variables. The results should 
depend on $q$ but we have not indicated this dependence.

Usual arguments of statistical mechanics imply that the following limit is well defined
\be
e(\alpha)=\lim_{N\to\infty}e(\alpha,N)
\ee
and that sample to sample fluctuations vanish when $N$ goes to infinity:
\be
\frac{\overline{\left(H^{*}\right)^{2}}-\left(\overline{H^{*}}\right)^{2}}{N^{2}} \to 0
\ee

According to the previous discussion there must be a phase transition at a critical value of 
$\alpha_{c}$, such that
\bea
e(\alpha)=0  \for \alpha<=\alpha_{c}\ , \\
e(\alpha)>0  \for \alpha>\alpha_{c} \ .
\eea
A simple argument shows that $e(\alpha)$ is a continuos differentiable function, so that the 
satisfiability-unsatisfiability transition cannot be a first order transition in the thermodynamic sense.

The goal of the statistical mechanics approach is to compute $\alpha_{c}$ and eventually $e(\alpha)$.
We would like also to compute the entropy density, that is related to the number of zero energy
configurations, but we will not address this point in this lecture.  We will see that in order to
compute $\alpha_{c}$ we will need a generalization of the entropy, i.e. the complexity
$\Sigma(\alpha)$ that is the exact equivalent for this problem of the configurational entropy used in
the study of glasses.

This random inequality model has never been studied, as far as I know, however the computation of
$\alpha_{c}$ should be relatively straightforward, for not too large values of $q$.  We will consider
in the next section a much more studied problem: the coloring of a random graph with $q$ different
colors.

\section{An intermezzo on random graphs}

The definition of random graphs have been discussed in Havlin's lectures, however it is convenient to recall here the main 
properties \cite{Erdos_Renyi}. 

There are many variants of random graphs: fixed local coordination number, Poisson distributed local
coordination number, bipartite graphs\ldots They have the same main topological structures in the
limit where the number ($N$) of nodes goes to infinity.

We start by defining the random Poisson graph in the following way: given $N$ nodes we consider the
ensemble of \emph{all} possible graphs with $M=\alpha N$ edges (or links).  A random Poisson graph is a
generic element of this ensemble.
   
The first quantity we can consider for a given graph is the local coordination number $z_{i}$, i.e.
the number of nodes that are connected to the node $i$.  The average coordination number $z$ is the
average over the graph of the $z_{i}$:
\be
z=\frac{\sum_{i=1,N}z_{i}}{N}
\ee
In this case it is evident that
\be
z=2 \alpha
\ee
It takes a little more work to show that in the thermodynamic limit ($N \to \infty$), the probability 
distribution of the local coordination number is a Poisson distribution with average $z$.

In a similar construction two random points $i$ and $k$  are connected with a
probability that is equal to $z/(N-1)$. Here it is trivial to show that the probability
distribution of the $z_{i}$ is Poisson, with average $z$. The total number of links is just
$zN/2$, apart  from corrections proportional to $\sqrt{N}$. The two Poisson ensembles, i.e. fixed
total number of links and fluctuating total number of links, cannot be distinguished locally for
large $N$ and most of the properties are  the same.

Random lattices with fixed coordination number $z$ can be easily defined; the ensemble is just given by
all the graphs with $z_{i}=z$ and a random graph is just a generic element of this ensemble .

One of the most important facts about these graphs is that they are locally a tree, i.e. they are
locally cycleless.  In other words, if we take a generic point $i$ and we consider the subgraph
composed by those points that are at a distance less than $d$ on the graph \footnote{The distance
between two nodes $i$ and $k$ is the minimum number of links that we have to traverse in going from
$i$ to $k$.}, this subgraph is a tree with probability one when $N$ goes to infinity at fixed $d$.
For finite $N$ this probability is very near to 1 as soon as
\be
\ln (N) > A(z)\; d \ ,
\ee
$A(z)$ being an appropriate
function. For large $N$ this probability is given by $1-O(1/N)$.

If $z>1$ the nodes percolate and  a finite fraction of the graph belongs to a single giant connected
component.  Cycles (or loops) do exist on this graph, but they have typically a length proportional
to $\ln(N)$.  Also the diameter of the graph, i.e. the maximum distance between  two points of the
same connected component is proportional to $\ln(N)$.  The absence of small loops is crucial because
we can study the problem locally on a tree and we have eventually to take care of the large loops
(that cannot be seen locally) in a self-consistent way.  i.e. as a boundary conditions at infinity.
This problem will be studied explicitly in the next section for the ferromagnetic Ising model.
\subsection{The Bethe Approximation}

Random graphs are sometimes called Bethe lattices, because a spin model on such a graph can be
solved exactly using the Bethe approximation.
Let us recall the Bethe approximation for the two dimensional Ising model. 

In the standard mean field approximation, one writes a variational principle assuming the all the
spins are not correlated \cite{I}; at the end of the computational one finds that the magnetization satisfies
the well known equation
\be
m=\tgh( \beta J z m)
\ee
where $z=4$ on a square lattice ($z=2d$ in $d$ dimensions) and $J$ is the spin coupling ($J>0$ for a
ferromagnetic model).  This well studied equation predicts that the critical point (i.e. the point
where the magnetization vanishes) is $\beta_{c} =1/z$.  This result is not very exiting in two
dimensions (where $\beta_{c}\approx .44$)  and it is very bad in one dimensions (where
$\beta_{c}=\infty$).  On the other end it becomes more and more correct when $d \to \infty$.

A  better approximation can be obtained if we look to the system locally and we compute the
magnetization of a given spin ($\sigma$) as function of the magnetization of the nearby spins
($\tau_{i}$,  $i=1,4$). If we assume that the spins $\tau$ are uncorrelated, but have magnetization
$m$, we obtain that the magnetization of the spin $\sigma$ (let us call it $m_{0}$) is given by:
\be
m_{0}= \sum_{\tau} P_{m}[\tau] \tgh(\beta J \sum_{i=1,4}\tau_{1}) \ ,
\ee
where 
\be
P_{m}[\tau]= \prod_{i=1,4} P_{m}(\tau_{i}), \ \ \ \ \ \  
P_{m}(\tau)=\frac{1+m}{2}\delta_{\tau,1}+\frac{1-m}{2}\delta_{\tau,-1} \ .
\ee
The sum over all the $2^{4}$ possible values of the $\tau$ can be easily done. 

If we impose the
self-consistent condition
\be
m_{0}(m)=m \ ,
\ee
we find an equation that enables us to compute the value of the magnetization $m$.

This approximation remains unnamed (as far as I know) because with a little more work we can get
the better and simpler Bethe approximation. The drawback of the previous approximation is that the 
spins $\tau$ cannot be uncorrelated because they interact with the same spin $\sigma$: the effect
of this correlation can be taken into account ant this leads to the Bethe approximation.

Let us consider the system where the spin $\sigma$ has been removed. There is a cavity in the
system and the spins $\tau$ are on the border of this cavity. We assume that in this situation these 
spins are uncorrelated and  they have a magnetization $m_{C}$. 
When we add the spin $\sigma$,
we find that the probability distribution of this spin is proportional to
\be
\sum_{\tau} P_{m_{C}}[\tau]) \exp\left(\beta J \sigma \sum_{i=1,4}\tau_{i}\right) \ .
\ee
The magnetization of the spin $\sigma$ can be computed and after some simple algebra we get
\be
m=\tgh\{z \;\arth[ \tgh(\beta J) m_{C}] \}\ , \label{BETHE}
\ee
with $z=4$.

This seems to be a minor progress because we do not know $m_{C}$. However we are very near the
final result. We can remove one of the spin $\tau_{i}$ and form a larger cavity (two spins removed).
If in the same vein we assume that the spins on the border of the cavity are uncorrelated and they
have the same magnetization $m_{C}$, we obtain
\be
m_{C}=\tgh\{(z-1) \arth[ \tgh(\beta J) m_{C}] \}\ . \label{CAVITY}
\ee
Solving this last equation we can find the value of $m_{C}$ and using the previous equation we can 
find the value of $m$.
                    
It is rather satisfactory that in 1 dimensions ($z=2$) the cavity equations become
\be
m_{C}=\tgh(\beta J) m_{C}\ .
\ee
This equation for finite $\beta$ has no non-zero solutions, as it should be.

The internal energy can be computed in a similar way: we get that the energy density per link is
given by
\be
E_{link}={ \tgh(\beta J) +m_{C}^{2}\over 1+ \tgh(\beta J)m_{C}^{2}}
\ee
and we can obtain the free energy by integrating the internal energy as function of $\beta$.

In a more sophisticated treatment we write the free energy as function of $m_{C}$:
\be
{\beta F(m_{C}) \over N}= F_{site}(m_{C})- \frac{z}{2} F_{link}(m_{C}) \ , \\
\ee
where $F_{link}(m_{C})$ and 
$F_{site}(m_{C})$ are appropriate functions \cite{HH}.
This free energy is variational, in other words the equation
\be
{\partial F \over \partial m_{C}} =0
\ee
coincides with the cavity equation (\ref{CAVITY}).
However for lack of space we will not discuss this interesting  approach \cite{HH,MP1}.

\subsection{Bethe lattices and replica symmetry breaking}

It should be now clear why the Bethe approximation is correct for random lattices. 
If we remove a
node of a random lattice, the nearby nodes (that were at distance 2 before) are now at a very large
distance, i.e. $O(\ln(N))$. In this case we can write
\be
\lan \tau_{i_{1}}\tau_{i_{2}} \ran \approx m_{i_{1}}m_{i_{2}}
\ee
and everything seems easy. 

This is actually easy in the ferromagnetic case where in absence of magnetic field at low temperature
the magnetization may take only two values ($\pm m$). In more complex cases, (e.g.
antiferromagnets) there are many different possible values of the magnetization because there are
many equilibrium states and everything become complex (as it should) because the cavity equations
become equations for the probability distribution of the magnetizations \cite{MPV}. 

This case have been long studied in the literature and for historical reasons it is usually said that
the replica symmetry is spontaneously broken \cite{MPV,PBOOK}.  Fortunately for the aims of this
lecture we need only a very simple form of replica symmetry breaking (we are be interested to the
zero temperature case) and we are not going to describe the general formalism.

\section{Coloring a graph}
\subsection{Basic definitions}
For a given graph $G$ we would like to know if using $q$ colors the graph can be colored in such a
way that adjacent nodes have different colors. It is convenient to  introduce the Hamiltonian
\be
H_{G}=\sum_{i,k}A(i,k)\delta_{\sigma_{i},\sigma_{i}}\ ,
\ee
where $A_{G}(i,k)$ is the adjacency matrix (i.e. 1 if the two nodes are connected, 0 elsewhere) and
the variable $\sigma$ may take values that go from 1 to $q$.  This Hamiltonian describes the
antiferromagnetic Potts model with $q$ states.  The graph $G$ is colourable if and only if the ground
state of this Hamiltonian ($E_{G}$) is zero.

For large $N$ on a random graph we expect that energy density
\be
{E_{G}\over N}=e(z)
\ee
does not depend on $G$: it should depends only on 
 the average coordination number $z$ of the graph.
It can be proved that
\bea
e(z) =0 \for z<1\ , \\
e(z) \propto \sqrt{z} \for z \to \infty \ .
\eea
Consequently there must be a phase transition at $z_{c}$ between the colorable phase $e(z)=0$ and the
uncolorable phase $e(z)\ne 0$.  Although it is possible to compute the function $e(z)$ for all $z$ in
this lectures we address only to the simpler problem of computing the value of $z_{c}$
\cite{MPWZ}.

For $q=2$ $z_{c}=1$, as it can be intuitively understood: odd loops cannot be coloured and large for
$z>1$ there are many large loops that are even and odd with equal probability.  The $q=2$ case is an
antiferromagnetic Ising model on a random graph, i.e. a standard spin glass \cite{E,MP1}.

\subsection{The cavity equations}

Let us start with the basic definitions.  Let us consider a legal coloring (i.e all adjacent nodes
have different colors).  We take a node $i$ and we consider the subgraph of nodes at distance $d$
from a given node.  Let us call $\cB(i,d)$ the interior of this graph.  With probability one (when
$N$ goes to infinity) this graph is a tree and the nodes at distance less or equal to $d$ are the
leafs of this tree (there may be also other leafs at shorter distance).  In the future we shall
assume that this graph is a tree and all the statements we shall do will be valid only with
probability one when $N$ goes to infinity.

We ask the following questions: 
\begin{itemize}
\item Are there other legal colorings of the graph that coincide with the original coloring outside
$\cB(i,d)$ and differs inside $\cB(i,d)$?  (Let us call the set of all these coloring  $\cC(i,d)$.)
\item
Which is the list of colors that the node $i$ may have in one of the coloring belonging to 
$\cC(i,d)$? (Let us call this list $L(i,d)$. This list depends on the legal configuration $\sigma$,
however for lightening the notation we will not indicate this dependence in this section.)
\end{itemize}

The cardinality of $L(i,d)$ increases with $d$ so that $L(i,d)$ must have a limit when $d$ goes to
infinity (still remaining in the region $d<<A(z) \ln(N)$).  We call this limit $L(i)$.  We have not
to forget that $L(i)$ is fuzzy for finite $N$, but it becomes sharper and sharper when $N$ goes to
infinity.  In other words $L(i)$ is the list of all the possible colors that the site $i$ may have if
we change only the colors of the nearby nodes and we do not change the colors of faraway nodes
\cite{P1,P5}.

Let us study what happens on a graph where the site $i$ has been removed.  We denote by $k$ a node
adjacent to $i$ and we call $L(k;i,d)$ the list of the possible colors of the node $k$ while the
colors outside $\cB(i,d)$ are fixed.  The various nodes $k$ do not interact directly and in this
situation they can have a color independently from the color of the other nodes.

In this situation it is evident  that $L(i,d+1)$ can be written as function of all the $L(k;i,d)$.
In order to write explicit equations it is convenient to  indicate by an overbar 
($\ba{L}$)  the list of the forbidden colors \footnote{The list of forbidden colors is just the list 
of colors that do not belong to the list of possible colors. From the point of view of set theory
the set of forbidden color is just the complement of the set of possible colors. Using this
notation we obviously have $\ba{\ba{L}}=L$.}. Let us us indicate by 
$F(L)$ the list of colors that must be forbidden at the nodes that are adjacent to a node where the
list of allowed colors is $L$. Barring the case where the list $L$ is empty it is easy to obtain
that:
\begin{itemize}
	 \item $F(L)=L$ if $L$ contains only one element.
	 \item $F(L)=\emptyset$ ($\emptyset$ being the empty set) if $L$ contains more than one element.
\end{itemize}
With these preliminaries we have that 
\be
L(i,d+1)=\ba{ \bigcup_{k} F(L(k;i,d))}\ ,
\ee
where $k$ runs over all the nodes adjacent to $i$, and the union is done in a set theoretical sense.

The previous formula can be easily transcribed into words. We have to consider all the neighbours
($k$)
of the node $i$; if a neighbour may be colored in two ways, it  imposes no constraint, if it can be
colored in only one way, it forbids the node $i$ to have its color. Considering all nearby nodes we
construct the list of the forbidden colors and the list of the allowed colors is just composed by
those colors that are not forbidden.

The previous formula is exact (with probability one as
usual). Also the next formula is valid  with probabilism one:
\be
L(i,d)=\ba{ \bigcup_{k} F(L(k;i,d))}\ ,
\ee
because in most of the cases $L(i,d+1)=L(i,d)$.

If we do the limit $d \to \infty  $ in any of the two previous formulae we get
\be
L(i)=\ba{ \bigcup_{k} F(L(k;i))}\ .
\ee
Of course everything is true with probability 1 when $N \to \infty$.

A further simplification in the previous formulae may be obtained if we associate to a list $L$ a
 variable
$\omega$ that take values from 0 to $q$ defined as follow
\begin{itemize}
	 \item The variable $\omega$ is equal to $i$ if the list contains only the $i^{\mbox{th}}$ color.
	 \item The variable $\omega$ is equal to 0 if the list contains more than one color.
\end{itemize}
Let us call $\Omega(L)$ this mapping. 

In the nutshell we have introduced an extra color, white, and we say 
that a site is white if it can be colored in more than two ways without changing the colors of the 
far away sites \cite{P1,P5,BMWZ}.
The rational for introducing the variable $\omega$ is that $F(L)$ depends only on $\Omega(L)$.
The previous equations induces equations for the variables $\omega$, i.e.
\be 
\omega(i)=\Omega\left(\ba{ \bigcup_{k} F(\omega(k;i))}\right)
\ee
In the same way as before we have to consider all the neighbours
($k$)
of the node $i$; if a neighbour is white, it  imposes no constraint, if it is colored
it forbids the node $i$ to have its color. Considering all nearby nodes we
construct the list of the forbidden colors. If more than one color is \emph{not} forbidden,
 the node is white, if only one color is \emph{not} forbidden, the node has this color.

The previous equation is just the generalization of eq.  (\ref{BETHE}) where we have the colors,
white included, instead of the magnetizations.  We have discrete, not continuos variables,
because we are interested in the ground state, not in the behaviour at finite temperature.  We have
to write down the equivalent of the cavity equations, eq.  (\ref{BETHE}), i.e. for the quantities
$\omega(i;l))$.  They are  given by
\be 
\omega(i;l)=\Omega\left(\ba{ \bigcup_{k \ne l} F(\omega(k;i))}\right) \ . \label{BELIEF}
\ee
We can also say that $\omega(i;l)$ is the information that is transmitted from the node $i$ to the
node $l$ and it is computed using the information transmitted by all the nearby nodes, with the
exclusion of $l$.

The previous equation are called the belief equations (sometimes the strong belief equations, in
order to distinguish them from the weak belief equations that are valid at non-zero temperature).  We
can associate to any legal coloring a solution (or a quasi-solution) of the belief equations in a
constructive way.  Sometimes the solution of the belief equations is called a whitening \cite{P1,P5},
because some nodes that where colored in the starting legal configuration becomes white.

The reader should notice that at this stage we can only say that the belief equations should be
satisfied in a fraction of nodes that goes to 1 when $N$ goes to infinity. However it is possible
that the total number of nodes where the beliefs equations are not satisfied remains finite or goes
to infinity with $N$ and for this reason in general we can only say that these equations have
quasi-solutions, not true solutions \cite{P1,P5}.

\section{Analytic computations}
\subsection{An interpretation}

The main reasons for introducing the whitening is that each legal coloring has many other coloring 
nearby that differs only by the change of the colors of a small number of nodes. The number of
these legal coloring that can be reached by a given coloring by making this kind of moves is
usually exponentially large. On the other hands   two colorings that differ only 
 by the change of the colors of a small number of nodes correspond to the same whitening.
 
Let us be more precise.  For each coloring let us write 
\be
\omega(i|\{\sigma\})=\omega(i,d|\{\sigma\}) \for 1<<d<<A(z)\ln(N) \ ,  
\ee 
where we have indicated in an explicit way that the whitening $\omega(i|\{\sigma\})$ depends on the
legal coloring $\{\sigma\}$.  

We say that two whitening are equivalent if they differs in a fraction of nodes less
that $\epsilon(N)$, where $\epsilon(N)$ is a function that goes to zero when $N \to \infty$.  For
finite $N$ everything is fuzzy and depends on the precise choice of $d$ and $\epsilon$. Let us
suppose that for large $N$ with probability one two whitening are equivalent or they differs in a
large fraction of nodes.

Generally speaking we have  three possibilities. 
\begin{itemize}
	 \item 
	 For all the legal configurations the corresponding whitenings ($\omega(\{\sigma\})$) are such that
	 $\omega(i,\{\sigma\})=0$, i.e. all nodes are white.
	 \item
	 For a generic  legal configurations the corresponding whitening is non-trivial. 
	 i.e. for a finite fraction of the nodes
	 $\omega(i,\{\sigma\})\!=0$.
	 \item The graph is not colorable and there are no legal configurations.
 \end{itemize}
 
 In the second case we would like to know how many whitenings are there, how they differs and which
 are their properties, e.g. how many sites are colored. 
 We shall see in the next section how these
 properties 
 may be computed analytically  and as byproduct we will find the value of $z$ ($z_{c}$) that
 separates the colorable phase from the uncolorable phase.

 In the case where there are many whitenings one can argue that the set of all the legal
 configurations breaks in an large number of different disconnected regions that are called with many
 different names in the physical literature \cite{MPZ,MP1,MP2} (states, valleys, clusters, lumps\ldots).
 Roughly speaking the set of all the legal configurations can be naturally decomposed into clusters
 of proximate configurations, while configurations belonging to different clusters (or regions) are
 not close.   The precise definition of
 these regions is rather complex \cite{PARISILH}; roughly speaking we could say that two legal
 configurations belongs to the same region if they are in some sense adjacent, i.e. they belongs to a
 different region if their Hamming distance is greater than $\epsilon N$.  In this way the precise
 definition of these regions depends on $\epsilon$, however it can be argued that there is an
 interval in $\epsilon$ where the definition is non-trivial and is independent from the value of
 $\epsilon$ \footnote{For a rigorous definition of these regions see \cite{TALE,DuMa,CDMM}.}.
 It is usually assumed that each whitening  is associated to a
 different cluster of legal solutions.  
 
 \subsection{Surveys}
 
Let us consider the case (that we suppose to be non-empty) where there is a large number of
non-equivalent whitening and we want to study the properties of this ensemble. At this end it is
convenient to introduce the probability that for a generic whitening $\omega$ of the ensemble we
have that $\omega(i)=c$; we will denote this probability $P_{i}(c)$.
We obviously have
\be
\sum_{c=0,q}P_{i}(c)=1 \ .
\ee

The quantities $P_{i}(c)$ generalize the physical concept of magnetization. In the statistical
analysis of the ensemble of configurations of an Ising model, the local variables may have only
two values ($\pm 1$) and 
\be
P(\pm 1)= {1\pm m_{i} \over 2} \ ,
\ee
where $m_{i}$ is the magnetization at the site $i$. Here $q+1$ colors are possible (white included)
and the magnetization is a $q+1$-dimensional vector, normalized to 1.

In order to do some computation we have to consider the also the two colors probabilities
$P_{i,l}(c_{1},c_{2})$.  We will assume a factorization hypothesis: for points $i$ and $l$ that are
far away on the graph the probability $P_{i,l}(c_{1},c_{2})$ factorizes into the product of two
independent probabilities
\be
P_{i,l}(c_{1},c_{2})=P_{i}(c_{1})P_{l}(c_{2}) \ ,
\ee
neglecting corrections that go to zero (in probability) when $N$ goes to infinity.
This hypothesis in not innocent: there are many case where it is not correct; however we cannot
discuss here this important and subtle point.
In order to do some computation we have to consider the also the two colors probabilities
$P_{i,l}(c_{1},c_{2})$.  We will assume a factorization hypothesis: for points $i$ and $l$ that are
far away on the graph the probability $P_{i,l}(c_{1},c_{2})$ factorizes into the product of two
independent probabilities \be P_{i,l}(c_{1},c_{2})=P_{i}(c_{1})P_{l}(c_{2}) \ee neglecting
corrections that go to zero (in probability) when $N$ goes to infinity.  This hypothesis in not
innocent: there are many case where it is not correct; however we cannot discuss here this important
and subtle point.

A similar construction can be done with the cavity coloring and in this way we define the
probabilities $P_{i;k}(c)$, where $k$ is a neighbour of $i$.  These probabilities are called surveys
(and they are denoted by $\vec{s}(i;k)$) because they quantifies the probability distribution of the
messages sent from the node $i$ to the node $k$.

Under the previous hypothesis the survey satisfy  simple equations. Let us see one of them, e.g
the one that relates $P_{i}(c)$ to the $P_{k;i}(c)$. The final equations are simple, but the
notation may becomes easily heavy, so let us write everything in an explicit way in the case where
the point $i$ has three neighbours $k_{1},k_{2},k_{3}$. The generalization is intuitive.

The formulae of the previous section tell us which  is the color of the site $i$ if we know the
colors $c_{1}=\omega(k_{1};i)$, $c_{2}=\omega(k_{2};i)$ and $c_{3}=\omega(k_{3};i)$. This relation 
can be written as
\be
c=T_{3}(c_{1},c_{2},c_{3})
\ee
Therefore the probability distribution of $c$ can be written as function of the probability
distribution of the $c_{i}$, that are supposed to factorize. We thus get
\be
P_{i}(c)=\sum_{c_{1},c_{2},c_{3}} P_{k_{1};i}(c_{1})P_{k_{2};i}(c_{2})P_{k_{3};i}(c_{3})
\delta_{c,T_{3}(c_{1},c_{2},c_{3})} \ .
\ee
Similar formulae can be written for computing the  surveys $P_{k_{1};i}$ as function of other
surveys, e.g.
\be
P_{i;k_{1}}(c)=\sum_{c_{2},c_{3}}P_{k_{2};i}(c_{2})P_{k_{3};i}(c_{3}) \label{SURVEY}
\delta_{c,T_{2}(c_{2},c_{3})} \ . 
\ee
In this way we obtain the so called survey propagation equations \cite{MPZ,MZ,P2}.

The survey propagation equations always have a trivial solution corresponding to all sites white:
$P_{i}(0)=1$ for all $i$.  Depending on the graph there can be also non-trivial solutions of the
survey equations.  Let us assume that if such a solution exist, it is unique (also this point should
be checked).  In the next section we will find the statistical properties of this solution and we
will
identify the values of $z$ where such a solution is present.

\subsection{An high level statistical analysis}

We are near the end of our trip.  Let us look to equation eq.(\ref{SURVEY}). The quantity
$P_{i;k_{1}}$ depends on $P_{k_{2};i}$ and $P_{k_{3};i}$: we indicate this relation by 
$P_{i;k_{1}}=\cT_{2}[P_{k_{2};i},P_{k_{3};i}]$.
However if we neglect the large loops the
quantities
 $P_{k_{2};i}$ and $P_{k_{3};i}$ do \emph{not} depend from $P_{i;k_{1}}$. 
 
If we consider the whole graph we can define the probability $\cP[P]$, i.e. the probability that a
given node has a probability $P(c)$.  If we consider only points with two neighbours we have
 \be
 \cP[P]=\int d\cP[P_{1}]d\cP[P_{2}] \delta\left[P-\cT_{2}(P_{1},P_{2}\right]
 \ee
If we sum over all the possible coordination numbers with the Poisson distribution the final
equation is
 \be
 \cP[P]=\sum_{n}{z^{n}\over n!} \int d\cP[P_{1}]\ldots d\cP[P_{n}] \delta[P-\cT_{n}(P_{1} \ldots
 P_{n})]
 \ee
 
We arrived to an  integral equation for the probabilities of the local probabilities (i.e. the
surveys).  This integral equation looks formidable and it is unlikely that it has an analytic
solution.  Fortunately often the solution of integral equations con be computed numerically without
too much difficulties on present days computers and this is what happens in this case.
 
One finds that there is a range $z_{d}<z<z_{U}$ where the previous integral equation has a
non-trivial solution and its properties can be computed \cite{P2}
The fact that survey equation has a non-trivial solution does not imply that there are
whitening that correspond to legal configurations so that at this stage we cannot compute the
critical value of $z$.  This problem will be solved in the next section.
 
 \subsection{The complexity}
 
In the same way that the entropy counts the number of legal colorings, the complexity counts the
number of different whitening; more precisely for a given graph we write
 \be
 \# \mbox{whitenings}= \exp (\Sigma_{G}) \ ,
 \ee
 where $\Sigma_{G}$ is the complexity.
 
 We assume that for large $N$ all graphs with the same average coordination number ($z$) have the
 same complexity density:
 \be
 \Sigma_{G}\approx N\Sigma(z) \ .
 \ee

\begin{figure}
 \begin{center}
 \includegraphics[angle=0,width=0.7\columnwidth]{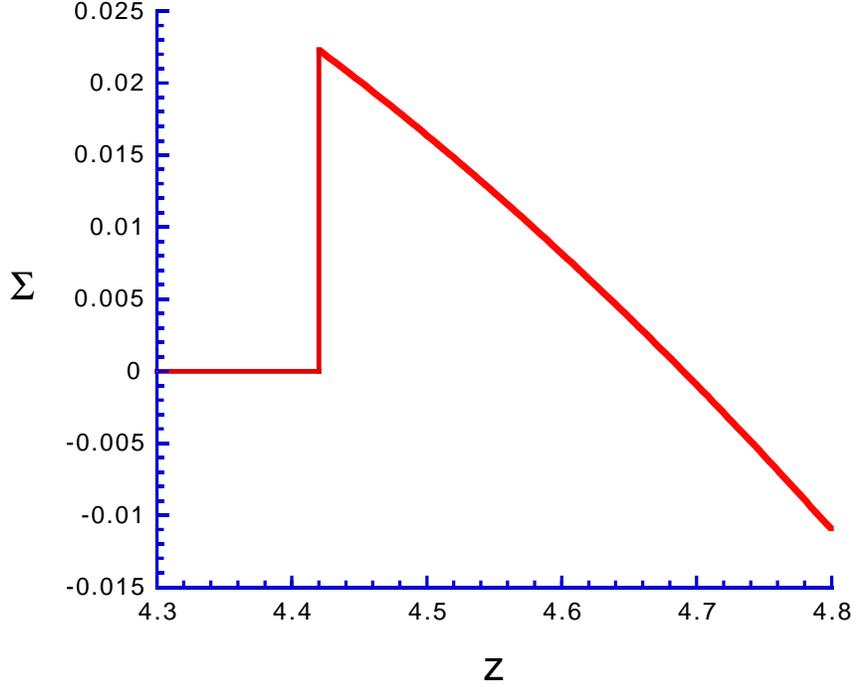}
 \caption{The complexity versus the average connectivity $z$ for $q=3$ (from \cite{MPWZ}).}
 \label{fig_eta}
 \end{center}
 \end{figure}

There is a simple way to compute the complexity.  It consists in counting the variation in the number
of whitenings when we modify the graph.  At this end it is convenient to consider the complexity as
function of $N$ and of $M$ (i.e. the total number of edges).  Asymptotically we should have
 \be
 \Sigma(M,N)= N\Sigma\left(\frac{2M}{N}\right)\ .
 \ee
There are two possible modifications we shall consider: adding one edge or adding one site (with the
related edges).

Let us suppose that we add an edge between the nodes $i$ and $k$.  Only those whitening in which the
colors of two nodes is different may be extended to the larger graph.  We thus find
 \bea
 \# \mbox{whitenings}(N,M+1)=
 \# \mbox{whitenings}(N,M)\left(1-\sum_{c=1,q}\omega_i(c)\omega_k(c)\right) \\
 \Sigma (N,M+1)=\Sigma (N,M)+ 
 \ln\left(1-\sum_{c=1,q}\omega_i(c)\omega_k(c)\right)\equiv \Sigma (N,M)+\Delta
 \Sigma_{\mbox{edge}}(z)
 \eea
 
In the same way if we add a site (and automatically $z$ links in the average) we get that the
complexity increases by an addictive factor $\Delta \Sigma_{\mbox{site}}$ that can be easily computed
(it is equal to the logarithm of the probability that the total number of different non-white colors
of the nodes to which the new added node is linked is less that $q$).
 
However if we want to change $N$ by one unit at fixed $z$ we have to add one site and only $z/2$ edges.
Putting everything together we find that
\be
\Sigma(z)=\Delta\Sigma_{\mbox{site}}(z)-\frac{z}{2} \Delta\Sigma_{\mbox{edge}}(z)
\ee

We know the probability distribution of the variables $\omega$ from the previous analysis. We can
now get the results for the complexity. It is shown in figure (1) for the case of the three
coloring. 

The complexity jumps from 0 to a finite value at $z_{d}=4.42$ and it decreases with increasing $z$
and becomes eventually negative at $z=4.69$.  A negative value of $\Sigma$ implies a number of
whitenings less than one and it is interpreted as the signal there there are no whitening (and no
legal configurations) with probability 1.  Indeed an explicit computation shows that in the region
where the complexity is negative a correct computation of the energy $e(z)$ gives a non-zero
(positive) result.  The value where the complexity becomes zero is thus identified as the
colorability threshold $z_{c}=4.69$.  We have thus obtained \cite{MPWZ} the result we wanted for
$q=3$.  Similar results may be obtained for higher values of $q$ \cite{MPWZ}.

\section{Conclusions}

The methods we have used in this lectures have been developed in the study of disordered systems,
(i.e. spin glasses) firstly  in the context of infinite range models (i.e. in the
limit where $z \to \infty$).  They have been extended to finite range models first at finite
temperature  and later at zero temperature (where notable simplifications are present).
The concept of complexity emerged in the study of the spin glasses and it was also introduced in
the study of glasses under the name of configurational entropy. The behaviour of the complexity as 
function of $z$ is very similar to what is supposed to happen in glasses as function of $\beta$.

These methods have been successfully applied to combinatorial optimization in the case of the $K$
satisfiability where our goal is to find a set of $N$ boolean variables that satisfy $\alpha N$
logical clauses (of the OR type) containing $K$ variables.  Also in this case the satisfiability
threshold \cite{KS,01} can be computed (for example we found \cite{MPZ} that for $K=3$
$\alpha_{c}=4.267$).

The research on these subjects is quite active. There are many problems that are open.
\begin{itemize}
	 \item The extension to other models.  This is interesting per se; moreover surprises may be
	 present.  
	 \item Verification of the self-consistency of the different hypothesis done and the
	 identification of effective strategies in the case where they fail \cite{MoRi2,MPR}.  
	 \item The construction of effective algorithms for finding a solution of the optimization
	 problem.  A first algorithm has been proposed \cite{MPZ,MZ,P3} and it has been later improved by
	 adding a simple form of backtracking \cite{PB}.  A goal is to produce an algorithm that for large
	 $N$ finds a solution with probability one on a random graph in a polynomial time as soon as
	 $z<z_{c}$ (i.e. in a computer time that is bounded by $N^{\lambda(z)}$).  Finding this algorithm
	 is interesting from the theoretical point of view (it is not clear at all if such an algorithm
	 does exist) and it may have practical applications.
	 \item
	 Last but not the least everything that we said up to now was derived using physicists stile.  One
	 should be able to transform the results derived in this way into rigorous theorems.  It is very
	 interesting that after a very long effort Talagrand \cite{Tala}, using some crucial results of
	 Guerra \cite{GUERRA}, has been recently able to prove that a similar, but more complex,
	 construction gives the correct results in the case of infinite range spin glasses (i.e. the
	 Sherrington Kirkpatrick model) that was the starting point of the whole approach.  It is
	 relieving to know that the foundations of this elaborate building are sound.  Some rigorous
	 results have also been obtained for models with $z$ \cite{FraLeo}.
\end{itemize}

The whole field is developing very fast and it is likely that in a few years we should have a
much more clear and extended picture.

\end{document}